\documentstyle[prb,aps,tighten,eqsecnum,floats,epsf]{revtex} 
\preprint{T98-134}
\newcommand{\nls}{NL$\sigma$M }

\begin{document}
\draft
\title{Generalized nonlinear sigma model approach to alternating 
spin chains and ladders}
\author{M. Bocquet and Th. Jolic\oe ur\thanks{e-mail: mbocquet, 
thierry@spht.saclay.cea.fr}}
\address{Service de Physique Th\'eorique, CEA Saclay, F91191 
Gif-sur-Yvette, France}
\date{March 1999}
\maketitle
%%%%%%%%%%%%%%%%%%%%%%%%%%%%%%%%%%%%%%%%%%%%%%%%%%%%%%%%%%%%%%%%% 
\begin{abstract}
We generalize the nonlinear sigma model treatment of quantum spin 
chains to cases including ferromagnetic bonds. When these bonds are
strong enough, the classical ground state is no longer the
standard N\'eel order and we present an extension of the known 
formalism to deal with this situation. We study the alternating 
ferromagnetic-antiferromagnetic spin chain introduced by Hida.
The smooth crossover between decoupled dimers and the Haldane
phase is semi-quantitatively reproduced. We study also a spin
ladder with diagonal exchange couplings that interpolates
between the gapped phase of the two-leg spin ladder
and the Haldane phase. Here again we show that there is a
good agreement between DMRG data
and our analytical results.
\end{abstract}
%%%%%%%%%%%%%%%%%%%%%%%%%%%%%%%%%%%%%%%%%%%%%%%%%%%%%%%%%%%%%%%%% 
%%%%%%%%%%%%%%%%%%%%%%%%%%%%%%%%%%%%%%%%%%%%%%%%%%%%%%%%%%%%%%%%% 
\section{Introduction}
One-dimensional quantum spin systems exhibit remarkable physical 
properties. One of the most interesting case is the s=1 
antiferromagnetic Heisenberg spin chain. Contrary to the s=1/2 case, 
this system has a gap as predicted by Haldane\cite{FDM} and a finite 
spin 
correlation length. It is an example of a system which is disordered 
at zero temperature due to quantum fluctuations. The original 
conjecture has been checked experimentally\cite{exp}, 
numerically\cite{num}
as well as analytically\cite{schulz,AKLT,STN}. Although the ground 
state is disordered in the sense that spin correlations decay 
exponentially, there is a hidden topological order\cite{DNR} that is 
revealed in the bulk of the chain only by nonlocal observables or by 
ground state degeneracy in an open geometry. This hidden order is 
most clearly seen in the VBS wavefunction which is an approximate 
ground state of the spin-1 chain\cite{AKLT}. To construct this 
wavefunction one has first to write each spin s=1 as a triplet of two 
fictitious spins s=1/2. Then one couples  nearest-neighbor spins
 s=1/2 into singlets. This leads to a function which is 
obviously singlet and translation invariant. It is an excellent 
approximation of the true ground state\cite{nous}. The perfect 
crystalline 
pattern of singlets is the hidden order. K. Hida has given an 
appealing picture of the Haldane gap and the VBS ground state by 
considering an alternating ferro-antiferromagnetic s=1/2 
chain\cite{Hida1}. The Hamiltonian is given by (see figure 1)~: 
\begin{equation}
	{\mathcal H}= J_{AF}\sum_{n} {\bf S}_{2n}\cdot {\bf S}_{2n+1} 	
+J_{F} \sum_{n} {\bf S}_{2n+1}\cdot {\bf S}_{2n+2}. 	\label{faf}
\end{equation}
Here ${\bf S}_{i}$ are s=1/2 spin operators, $J_{AF}$ is {\it 
positive} and $J_{F}$ is {\it negative}. In what follows, we set 
$J_{AF}=1$ and $J_{F}=-\gamma$. The family of systems defined by 
Eq.(\ref{faf}) has simple limiting cases. For $\gamma =0$ we have a 
set of decoupled pairs of spins that have a trivial ground state~: 
all pairs are locked in singlets. When $\gamma\rightarrow\infty$, the 
s=1/2 are coupled by pairs into s=1 states and we get a chain of 
spins s=1. Hida has studied numerically the gap of the system as a 
function of $\gamma$. He has used Lanczos diagonalization techniques 
to evaluate the gap and he showed that there is no
phase transition as a function of $\gamma$. As a consequence, the 
Haldane gap of the limit $\gamma\rightarrow\infty$ is continuously 
connected to the trivial gap of the decoupled limit $\gamma =0$. The 
excitation spectrum also evolves smoothly. So the alternating chain 
offers a simple physical picture of both the Haldane gap and the 
hidden topological order. 

There is another approach to quantum spin chains which is the 
continuum field theory known as the nonlinear $\sigma$ model 
(NL$\sigma$M).
Introduced originally by Haldane\cite{FDM}, this field theory 
includes a topological term\cite{Aff} $\theta =2\pi s$ for a spin-s 
chain. 
While this term can be discarded when s is integer, it is responsible 
for masslessness when $\theta =\pi $ (mod $2\pi$), i.e. for 
half-integer spin chain.
This approach has been recently applied \cite{Sierra,Ita,sene} to 
spin ladders where there is also a parity effect which is given by 
the number of legs of the ladder. Indeed, spin ladders with even 
number of legs are generically gapped while odd-numbered ones are 
gapless. This effect can be explained in the NL$\sigma$M framework 
with a topological term which is $\theta =2\pi s\times n_{l}$ where 
$n_{l}$ is the number of legs. It is also of great interest to 
consider generalized spin ladders with various types of bond 
alternation and/or additional exchange couplings. For example, a spin 
ladder with a diagonal coupling interpolates smoothly\cite{White}
between the s=1 chain and the two-leg spin ladder. Previous 
investigations\cite{Takano,Koga,Fukui1,Fukui2} 
of these more general situations have
used the NL$\sigma$M but with the impossibility to treat cases 
including ferro bonds as in the ferro-antiferro chain introduced by 
Hida.

In this work, we present a generalized \nls that includes additional 
massive modes and we show that this model is able to reproduce the 
s=1 limit of alternating chain of Hida. We discuss the construction 
of this new effective theory in the path-integral framework although 
it could also be done in the hamiltonian formalism with identical 
results. 

In Sect. \ref{1} we treat Hida's chain case. We compute the spin-wave 
spectrum, motivate the \nls approach and compute an approximate spin 
gap formula for the whole range of $\gamma$. In sect. \ref{2}, we 
apply the same formalism to White's mapping from the spin-1 chain to 
the antiferromagnetic two-leg ladder. Sect. \ref{3} contains our 
conclusions.

%%%%%%%%%%%%%%%%%%%%%%%%%%%%%%%%%%%%%%%%%%%%%%%%%%%%%%%%%%%%%%%%% 
\section{Ferro-antiferromagnetic spin chain} 
\label{1}
%%%%%%%%%%%%%%%%%%%%%%%%%%%%%%%%%%%%%%%%%%%%%%%%%%%%%%%%%%%%%%%%% 

\subsection{Spin-wave spectrum}

A \nls for the spin chain is a non perturbative field theory which 
is built on the lowest energy modes of the classical theory of 
the chain.
We first study this 
alternating spin chain at the classical level. 
The equations of motion for the ferro-antiferromagnetic spin chain 
follow from $\frac{d}{dt}{\bf n}_i=i[{\cal H},{\bf n}_i]$ and 
are given by~:
\begin{equation}
\left\{
\begin{array}{l}
\frac{d}{dt}{\bf n}_{2i}=-{\bf n}_{2i}\times(-\gamma {\bf n}_{2i+1} 
+{\bf n}_{2i-1}), \\
\frac{d}{dt}{\bf n}_{2i+1}={\bf n}_{2i+1}\times(-\gamma {\bf n}_{2i} 
+{\bf n}_{2i+2}),
\end{array}
\right.
\end{equation}
where ${\bf n}_i$ is the classical spin vector at site $i$. Then we 
linearize those equations around the Neel order configuration ${\bf 
n}_{2i}=(-1)^is {\bf u}+{\bf \alpha}_i$ and ${\bf n}_{2i+1}=(-1)^is 
{\bf u}+{\bf \beta}_i$ where $\bf u$ is an arbitrary unitary vector 
and ${\bf \alpha}_i$ and ${\bf \beta}_i$ are orthogonal planar vector 
fields. Note that $\bf u$ breaks the natural O(3) symmetry of the 
hamiltonian. We represent both ${\bf \alpha}_i$ and ${\bf \beta}_i$ 
by complex fields $\xi_i$ and $\chi_i$.
The linearized equations of motion then reads~: 
\begin{equation}
\left\{
\begin{array}{l}
\frac{d}{dt}{\xi}_{n}=is(-1)^n(-\gamma 
\chi_n+\chi_{n-1}+(1+\gamma)\xi_n), \\ 
\frac{d}{dt}{\chi}_{n}=is(-1)^n(-\gamma 
\xi_n+\xi_{n+1}+(1+\gamma)\chi_n). \end{array}
\right.
\end{equation}
A plane wave Ansatz for those fields is~:
\begin{equation}
\left\{
\begin{array}{l}
\xi_n=e^{i(\omega+kn)}(a(k)+(-1)^n b(k)), \\ 
\chi_n=e^{i(\omega+kn)}(c(k)+(-1)^n d(k)). \end{array}
\right.
\end{equation}
The eigenstates $(a,b,c,d)(k)$ of energy $\omega$ and momentum $k$ 
are the eigenvectors of the following matrix~:
\begin{equation}
s.\left[
\begin{array}{cccc}
0	& 1+\gamma	& 0	& -\gamma -e^{-ik} \\
1+\gamma	& 0	& -\gamma+e^{-ik} & 0	\\
0	& -\gamma-e^{ik} & 0	& 1+\gamma	\\
-\gamma+e^{ik} & 0	& 1+\gamma	& 0	\\
\end{array}
\right] ,
\end{equation}
with eigenvalues $\omega(k)$. The energy
$\omega$ is a solution of its characteristic polynomial 
\begin{equation}
\left(\frac{\omega}{s}\right)^4-4\gamma(1+\gamma) 
\left(\frac{\omega}{s}\right)^2+4\gamma^2 \sin^2 k=0. 
\end{equation}
The dispersion relation is given by~:
\begin{equation}
\omega(k)=s\sqrt{2\gamma(1+\gamma)\pm
2\gamma\sqrt{(1+\gamma)^2-\sin^2 k}}.
\end{equation}
The two lowest positive excitations correspond to momenta $k_c=0$ and 
$k_c=\pi$. Linearizing the dispersion relation around $k_c$, we 
obtain for those modes the linear relation 
$\omega=s\sqrt{\frac{\gamma}{1+\gamma}}|k-k_c|$. Hence the velocity 
of those spin waves is $v=s\sqrt{\frac{\gamma}{1+\gamma}}$. They 
stand for the two Goldstone modes of the (classically) broken $O(3)$ 
symmetry. We will ground our \nls on them, assuming they are slowing 
varying modes, in an theory where the rotational symmetry is 
unbroken. We also reasonably assume that the massive modes do not 
interfere significantly with the Goldstone modes, their energy being 
of the order of $2s\sqrt{\gamma(1+\gamma)}$.

%%%%%%%%%%%%%%%%%%%%%%%%%%%%%%%%%%%%%%%%%%%%%%%%%%%%%%%%%%%%%%%%% 
\subsection{The \nls mapping}
Here we briefly recall the derivation of the \nls for an 
antiferromagnetic 
spin-s chain with $s$ even.
So consider a spin-s antiferromagnetic spin chain with s even, 
characterized by the hamiltonian ${\cal H}=\sum_i J_{AF}{\bf 
S}_i.{\bf S}_{i+1}$. This spin chain can be mapped onto a \nls. The 
mapping can be performed within the lagrangian formalism if one uses 
coherent states representation for the spin operators. The discrete 
action is made of the exchange interaction terms and of the Berry 
phases $W[{\bf n}_i]$ of the spin vector field ${\bf n}_i$. 
\begin{equation}
S=\int dt \sum_i Js^2 {\bf n}_i.{\bf n}_{i+1}+ \sum_i s.W[{\bf n}_i]. 
\end{equation}
The coherent state vectors ${\bf n}_i$ are expanded according to 
\begin{equation}
\left\{
\begin{array}{l}
{\bf n}_{2i}={\bf l}_{2i}-s\Phi_{2i} \\
{\bf n}_{2i+1}={\bf l}_{2i+1}+s\Phi_{2i+1}, \\ \end{array}
\right.
\end{equation}
where the vector field $\Phi_i$ (local staggered magnetization) is 
unitary and the field ${\bf l}_i$ satisfy ${\bf n}_i.{\bf l}_i=0$. 
Those fields can be hinted at from the study of modes in the 
semi-classical spin chain.
Now suppose ${\bf l}_i$ is of order of magnitude $a$, the lattice 
spacing. This representation of the spin chain allows for a 
continuous form for the action $S$, which depends on the fields ${\bf 
l}(x,t)$ and $\Phi(x,t)$. Because we supposed ${\bf l}$ is of 
magnitude $a$, the action is quadratic in ${\bf l}$ so that the field 
can be integrated out.
Now recall that the spin magnitude s is even. Then the Berry phases 
of the spin coherent states partially gather to form a topological 
term $\theta=2\pi s$, the effect of which is null in the action. The 
final effective action in the field $\Phi$ is~:
\begin{equation}
S=\int dtdx \frac{1}{2g}\left(\frac{1}{c}(\partial_t \Phi)^2 
-c(\partial_x\Phi)^2\right) ,
\end{equation}
where $g$ is the coupling constant of the \nls: $g=2/s$ and $c$ is 
the velocity $c=2J_{AF}s$.

%%%%%%%%%%%%%%%%%%%%%%%%%%%%%%%%%%%%%%%%%%%%%%%%%%%%%%%%%%%%%%%%% 

\subsection{Effective \nls}

Let us consider now the alternating spin chain. The main difficulty 
we encounter in trying to represent the Hida chain is the fact that 
it is inhomogeneous, so that writing an adequate continuous action 
out of the discrete hamiltonian is not a trivial task. Indeed, the 
fundamental microscopic structure is a block of two spins (call them 
$1$ and $2$, say).
It will turn out that it is better to choose pairs of spins naturally 
coupled by the ferromagnetic link, preferably to the 
antiferromagnetic one.
This choice corresponds anyway to Hida's idea of pairing those spins 
s=1/2 to make them an effective spin-1 in the limit where $\gamma$ 
goes to infinity (large ferromagnetic coupling).

The most rigorous way we could imagine to handle the continuous limit 
would be to introduce two coherent states vector fields, one for each 
of the two sites. Unfortunately, this would yield an intricate 
 action, due to the appearance of several equally contributing 
massive modes. In particular, the Berry phases contribution of those 
coherent states would not be any more easily recognizable as a 
topological invariant. 

Now consider a two-block structure, made of the four spins ${\bf 
S}_{2i}^1,{\bf S}_{2i}^2,{\bf S}_{2i+1}^1$, and ${\bf S}_{2i+1}^2$. 
The pairs ${\bf S}_{2i}^1,{\bf S}_{2i+1}^1$ and ${\bf S}_{2i}^2,{\bf 
S}_{2i+1}^2$ are appropriate candidates to generate two \nls models.
A first (incorrect) idea, sustained by our wish to go to the 
continuous limit, is to assume that they form the same \nls. This 
assumption would be correct in the limiting case $\gamma$ goes to 
infinity, but far too crude in any other case. 

To cure this, at least partially, we add one extra field $\Delta_i$ 
to the two semi-classical \nls slow modes ${\bf l}_i$ and $\Phi_i$. 
It represents small quantum fluctuations, remnants of massive modes, 
that may bring about effective corrections to the \nls action.

Accordingly, the coherent states fields are decomposed as~:
\begin{equation}
\left\{
\begin{array}{l}
{\bf n}_{2i}^1={\bf l}_{2i}-s\Phi_{2i}+a\Delta_{2i} \\ {\bf 
n}_{2i}^2={\bf l}_{2i}-s\Phi_{2i}-a\Delta_{2i} \\ {\bf 
n}_{2i+1}^1={\bf l}_{2i+1}+s\Phi_{2i+1}-a\Delta_{2i+1} \\ {\bf 
n}_{2i+1}^2={\bf l}_{2i+1}+s\Phi_{2i+1}+a\Delta_{2i+1} . \end{array}
\right.
\end{equation}

Here $s$ is the spin magnitude and $a$ is the lattice spacing. Note 
that this lattice spacing is of the length of a two-spin block. The 
amplitude of the quantum fluctuation is of the order of this lattice 
spacing. This assumption make the problem a tractable one, since we 
do not need to consider derivatives of the field $\Delta_i$ when 
expanding the action. We enforce it by setting $a$ as a prefactor of 
the fluctuating field. Note that the standard  momentum field 
${\bf l}$ is implicitly assumed to be of order $a$.

We intentionally chose only one fluctuation field, contrary to 
Senechal's scheme \cite{sene} where the coherent state fields are 
decomposed on as many possible independent fluctuation fields. We now 
justify the choice of this particular field by means of a path 
integral reasoning.

We suspect that some highly fluctuating paths are contributing in the 
action. Indeed, exchange couplings vary on the microscopic scale by a 
macroscopic amount, and do not behave smoothly with respect to the 
position. As a consequence some irregular paths might be 
energetically favorable. Yet a straightforward continuous limit of 
the action would not retain them since they are not spatially 
regular. That is why we should enforce some possible (non derivable) 
contributing fluctuation in the paths.
Now let us see why we chose this particular field $\Delta_i$. The 
fields ${\bf l}_i$ and $\phi_i$ parametrize the variation of the path 
between the two-block pattern $(2i,2i+1)$, spatially indexed by $i$. 
So we need one field to represent the variation inside the two-spin 
blocks. In the block $2i$, the coherent state vectors ${\bf 
n}_{2i}^1$ and ${\bf n}_{2i}^2$ differ by an amount of $2a\Delta_i$. 
Since the microscopic pattern is a two-spin block, we have then no 
choice than to make the coherent state vectors ${\bf n}_{2i+1}^1$ and 
${\bf n}_{2i+1}^2$ differ by an amount of $-2a\Delta_i$, because the 
variation on the scale of the two-spin block are already taken into 
account within the \nls fields. And this exhausts contributing 
infinitesimal fluctuations of the path.

The action of the spin chain we get through the coherent states 
representation is of the form
\begin{equation}
S=\int dt \sum_i s^2 {\bf n}_{2i}.{\bf n}_{2i+1}-\gamma \sum_i 
s^2{\bf n}_{2i+1}.{\bf n}_{2i+2}+ \sum_i s.W[{\bf n}_i].
\end{equation}
One can expand the coupling terms in the action, then goes to the 
continuous limit, which yields
\begin{equation}
S_c=\int dt\frac{dx}{a} \left[-4a^2(1+\gamma)\Delta^2-4l^2 
+4a^2s\partial_x\Phi.\Delta-a^2 s^2(\partial_x \Phi)^2 \right]. 
\end{equation}
We emphasize the fact that the ferromagnetic exchange terms are of 
course expanded from its aligned configuration contrary to the 
antiferromagnetic exchange terms which are expanded from the Neel 
order. That is the reason why NL$\sigma$ models derived for 
antiferromagnetic ladder or alternating spin chain 
\cite{Takano,Koga,Fukui1,Fukui2} cannot be straightforwardly applied 
to the present cases~: they are not built upon the same 
semiclassical configurations. Note that since the measure element 
$dx$ is a two-spin block, it is equal to $a$.
The Berry phases of the spins are of the form $\int dtdx \delta{\bf 
n}.{\bf n}\wedge \partial_t {\bf n}$, where $\delta{\bf n}$ is the 
spatial variation of the field ${\bf n}$, and give
\begin{equation}
S_b=\int dt\frac{dx}{a} s(4{\bf l}+2as\partial_x \Phi). \Phi\wedge 
\partial_t \Phi.
\end{equation}

Since we sum up the contributions for a double two-spin block, we 
must divide the whole sum by a factor $2$. We end up with 
\begin{equation}
S=\int dx dt \left[ -2(1+\gamma) \Delta^2-2{\bf 
l}^2-2s\Delta.\partial_x \Phi -\frac{1}{2}s^2(\partial_x \phi)^2 
\right] +\int dx dt \left[ s^2(\partial_x \Phi).\Phi\wedge\partial_t 
\Phi +2s{\bf l}.\Phi\wedge\partial_t \Phi \right], \end{equation}
where we have set $a=1$ for commodity.
We recover a spin-2s topological $\theta$-term with $\theta=4\pi s$. 
This has the consequence that the topological term does not 
contribute, so that the spin chain is likely to be gapped.

We next integrate over the fluctuation fields, that is to say $\bf l$ and 
$\Delta$. We then obtain the \nls action \begin{equation}
S=\int dtdx \left[ \frac{1}{2}(\partial_t \Phi)^2 -\frac{1}{2}s^2 
\frac{\gamma}{1+\gamma}(\partial_x \Phi)^2 \right]. \end{equation}
The standard parameters of this \nls are the coupling constant $g$ 
and the velocity $c$ given by
\begin{equation}
g=\frac{1}{s}\sqrt{\frac{1+\gamma}{\gamma}} \qquad \mbox{and} \qquad 
c=s \sqrt{\frac{\gamma}{1+\gamma}}. \end{equation}
Note that $c$ perfectly matches the velocity $v$ we found for the 
classical spin waves.
When we make $\gamma$ goes to infinity the coupling constant $g$ goes 
to $1/s$ and the velocity goes to $s$.
These are the expected parameters for the 2s-\nls. The Hida's chain is 
build up of spin one half so that for $s=1/2$ $g$ goes to $2$ and $c$ 
to $1/2$ in
units of the antiferromagnetic exchange coupling. Those are the 
parameters of a spin-1 chain of exchange $J_{AF}/4$. This is the 
limit found by Hida for the alternating chain. Let us recall why the 
magnitude of the exchange coupling should be so. Indeed in this limit 
the ferromagnetic pairs are in a triplet state. Because of rotation 
invariance, the limiting hamiltonian can be written as an effective 
spin-1 chain. The coupling can only be quadratic ${\bf S}_i.{\bf S}_j$ 
or quartic $({\bf S}_i.{\bf S}_j)^2$. The quartic term is excluded in 
this limit because it corresponds to second order perturbation theory 
in $\gamma^{-1}$. The prefactor in front of the spin-1 coupling can 
be determined with the help of the calculation of a single matrix 
element, which is easily done on the all-spin-up configuration and 
yields $J_{\tiny \mbox{eff}}=J_{AF}/4$. So that in this limit, our 
\nls is consistent with Hida's argument.

\subsection{Energy gap evaluation in the large-$N$ limit} 

To evaluate the spin gap, we compute the mass generated by the \nls 
in the limiting case of a large number of components for the field 
$\Phi$($N\rightarrow \infty$).
In that limit a closed expression can be obtained for it, thanks to 
the large N saddle-point approximation. So we made $\Phi$ a 
N-component field and rescaled it by a factor $\sqrt{N}$.
We then enforce the unitary constraint on $\Phi$ by means of a 
conjugate field $\lambda$, so that
the unconstrained partition function is
\begin{equation}
Z=\int D\Phi D \lambda e^{-S+i\int dxdt \lambda(\Phi^2-N)}. 
\end{equation}

The saddle point equation for the complete action can now be safely 
derived, and we may look for a constant solution for $\lambda$ that 
we will call $im^2$. In the limit of large $N$, $m^2$ is the mass 
generated by the \nls.

\begin{equation}
\int \frac{dk}{2\pi}\frac{d\omega}{2\pi} 
\frac{1}{\frac{c}{g}k^2+\frac{1}{gc}\omega^2+m^2}=1. \end{equation}
In the limit $\gamma$ goes to $0$, the chain is totally dimerized and 
the velocity $c$ goes to $0$. Because of this unusual feature, and 
also because we wish to derive results valid for a large range of 
$\gamma$, we will not resort to a radial cut-off in the euclidean 
space-time of the \nls.
Rather we will first integrate on frequencies, then integrate on the 
momenta, with a large-momenta cut-off $\Lambda$. So that instead of 
the usual radial cut-off, we integrate over a strip in the plane 
$(k,\omega)$ along the $\omega$-axis. The reason for this is that 
when $\gamma$ goes to 0, the prefactor $cg^{-1}$ of $k^2$ goes also 
to zero whereas the prefactor $(cg)^{-1}$ of $\omega^2$ remains 
constant so that large frequencies are more and more relevant and 
must not be cut off. In the process (which actually corresponds to 
the decoupling of dimers), we lost the space dimension of space-time. 

The first integration is over $\omega$ and gives \begin{equation}
g\int_{-\Lambda}^\Lambda \frac{dk}{4\pi} 
\frac{1}{\sqrt{k^2+\frac{g}{c}m^2}}=1.
\end{equation}
Then integrating over $k$ we can extract the mass of the \nls 
\begin{equation}
m=\sqrt{\frac{c}{g}}
\frac{\Lambda}{\sinh(2\pi s \sqrt{\frac{\gamma}{1+\gamma}})}. 
\end{equation}
In the standard derivation of the mass generated by the \nls (see for 
example \cite{Auerbach}) one would exchange the function hyperbolic 
sine for the function exponential.
Indeed, in order for this computation to make sense we must have 
$\sqrt{gc^{-1}}m << \Lambda$.
Since in the \nls related to the spin-2s chain, $\sqrt{gc^{-1}}$ is 
finite, the (then meaningless) hyperbolic sine function can be 
replaced with the exponential function. Yet for our alternating 
chain, it can't be done since in the limit $\gamma$ goes to $0$ the 
argument of the function vanishes. Hence to encompass the full range 
of $\gamma$ we must retain the hyperbolic sine function. 

Now we can evaluate the energy gap~: 
\begin{equation}
\Delta_s=\sqrt{gc}.m=\Lambda \frac{s \sqrt{\frac{\gamma}{1+\gamma}}} 
{\sinh(2\pi s \sqrt{\frac{\gamma}{1+\gamma})}}. 
\end{equation}

Whatever the spin magnitude $s$,
the gap of an antiferromagnetic pair of spins is equal to the gap 
between the triplet state and the singlet, i.e. $J_{AF}$. Hence when 
$\gamma \rightarrow 0$, we can determine that $\Delta$ goes to 
$\Lambda/(2\pi)$.
This statement allows us to determine the cut-off which appears to be 
$2\pi$ (actually $2\pi/a$ but we set $a=1$). We can then write~:
\begin{equation}
\Delta_s=\frac{2\pi s \sqrt{\frac{\gamma}{1+\gamma}}} {\sinh(2\pi s 
\sqrt{\frac{\gamma}{1+\gamma})}}. 
\end{equation}

We have obtained a {\em self-contained}, estimate of the chain gap 
for the spin magnitude $s$ . For the spin-1 chain we obtain 
$\Delta_1=\frac{4\pi}{\sinh \pi}$, which is $1.08$ far from the 
numerically \cite{Golinelli} known $0.41$. We expect a better result 
for the spin-2 chain, closer to the "large spin limit". We obtain 
$\Delta_2=\frac{8\pi}{\sinh 2\pi}$, which is $0.094$ fairly close to 
the numerically \cite{Jolicoeur} known value of $0.085$.
As for the correlation lengths, we obtain $\xi_1\sim 2$ to compare 
with the numerically known $\xi_1\sim 6$ whereas we obtained 
$\xi_2\sim 43$ to compare with the numerical value $\xi_2\sim 49$ .

The figures labeled 3 and 4 are drawings of the curves of the spin 
gap w.r.t. $-\gamma$ in the range $\gamma \in [0,1]$ and $-1/\gamma$ 
in the range $\gamma \in [1,\infty[$. We did so in order to compare 
our results to Hida's presentation of his numerical computation 
\cite{Hida1}. Our result agrees qualitatively on the whole range of 
$\gamma$ including the limiting case $\gamma$ goes to zero. 

To check that our \nls approach is still valid in the limit $\gamma$ 
goes to zero, we can also compute the correlation length. It can be 
read on the saddle point equation~:
$\xi=\sqrt{\frac{c}{g}}\frac{1}{m}$.
Then we get~:
\begin{equation}
\xi=\frac{1}{2\pi}\sinh\left( 2\pi s \sqrt{\frac{\gamma}{1+\gamma}} 
\right). 
\end{equation}
We check that the correlation length goes to zero like $s\sqrt 
\gamma$ when $\gamma \rightarrow 0$, which is expected since at this 
point the chain is made of decoupled dimers.

%%%%%%%%%%%%%%%%%%%%%%%%%%%%%%%%%%%%%%%%%%%%%%%%%%%%%%%%%%%%%%%%% 
\section{An alternating spin ladder}
\label{2}

In Ref.~(\cite{White}), S.~R. White introduced several mappings that 
interpolate smoothly between spin ladders and a spin-1 chain within 
the Haldane'spin gap phase.
Here we treat one of these mappings.
It consists in an antiferromagnetic spin ladder with an additional 
diagonal bond in every plaquette formed by legs and rungs (see figure 2).
The exchange coupling associated to this bond is ferromagnetic (we 
will denote it as $D$) and does not introduce any frustration in the 
ladder.
The hamiltonian is given by
\begin{equation}
{\cal H}=J\sum_{n;a=1,2} {\bf S}^a_{n} {\bf S}^a_{n+1} +K \sum_{n} 
{\bf S}^1_{n} {\bf S}^2_{n}
+D \sum_{n} {\bf S}^1_{n} {\bf S}^2_{n+1} \end{equation}
where all exchange couplings are chosen positive. When $D$ goes to 
$0$ we recover the usual antiferromagnetic spin ladder. Whereas in 
the limit $-D$ goes to infinity, the pairs of ferromagnetically 
bounded spin are in a triplet state.
The effective hamiltonian can then be expanded by rotational 
invariance in terms of spin-1 couplings. The effective 
antiferromagnetic coupling constant is then evaluated on any matrix 
element of the hamiltonian and gives $J_{\tiny \mbox{eff}}=(2J+K)/4$. 
So this limit corresponds to an antiferromagnetic spin-1 chain. 
Since in the process, we remain in the Haldane gapped phase, we may 
apply our scheme to obtain an estimate of the gap as a function 
of the diagonal coupling $D$. 

%%%%%%%%%%%%%%%%%%%%%%%%%%%%%%%%%%%%%%%%%%%%%%%%%%%%%%%%%%%%%%%%%%%%%
\subsection{Effective \nls}

The microscopic pattern of the ladder is composed of the four spins 
${\bf S}_{2i}^1,{\bf S}_{2i}^2, {\bf S}_{2i+1}^1$, and ${\bf 
S}_{2i+1}^2$ forming a square and, apart from the diagonal bond, 
three bonds: one on each leg, and the last one on one of the two 
rungs closing the square (see figure 2). As for the Hida's chain case 
calculations are done on a doubled cell.

Possible candidates to spin pairs forming a \nls are the 
nearest neighbours on the same leg. Pairs of site linked by a rung 
contribute to the same \nls. Then, we will need only one extra 
fluctuation field to describe of fluctuating paths inside the 
elementary cell, {\em before} going to the continuous limit. 

Accordingly the coherent states fields are decomposed as~:
\begin{equation}
\left\{
\begin{array}{l}
{\bf n}_{2i}^1={\bf l}_{2i}-s\Phi_{2i}+a\Delta_{2i} \\ {\bf 
n}_{2i}^2={\bf l}_{2i}-s\Phi_{2i}+a\Delta_{2i} \\ {\bf 
n}_{2i+1}^1={\bf l}_{2i+1}+s\Phi_{2i+1}-a\Delta_{2i+1} \\ {\bf 
n}_{2i+1}^2={\bf l}_{2i+1}+s\Phi_{2i+1}-a\Delta_{2i+1} \end{array}
\right.
\end{equation}

In the following we set $J=1$, $K=\rho$ and $D=-\delta$ so that the 
coupling constants are expressed in units of the longitudinal 
antiferromagnetic coupling $J$.
One can then expand the coupling terms in the action, which yields~:
\begin{equation}
S_c=\int dt\frac{dx}{a} 
\left[-4a^2(\rho+\delta)\Delta^2-4(2+\rho){\bf l}^2 +4a^2\delta 
s\partial_x\Phi.\Delta-a^2 (\rho+\delta) s^2(\partial_x \Phi)^2 
\right] 
\end{equation}
with the same care for the ferromagnetic bonds as was done 
previously. Like for the Hida's chain the measure element $dx$ is a 
two-spin block and is equal to $a$. The Berry phases of the spins are 
of the form $\int dtdx \delta{\bf n}.{\bf n}\wedge \partial_t {\bf 
n}$ and give
\begin{equation}
S_b=\int dt\frac{dx}{a} 2 s {\bf l}.\Phi\wedge \partial_t \Phi. 
\end{equation}

Since we sum up the contributions for a double two-spin block, we 
must divide the whole sum by $2$. We end up with~:
\begin{equation}
S=\int dx dt \left[ -2(2+\rho){\bf 
l}^2-2(\rho+\delta)\Delta^2-2s\Delta. \partial_x 
\Phi-\frac{1}{2}s^2(2+\delta)(\partial_x \Phi)^2 \right] +\int dxdt 
\left[ 2{\bf l}s\Phi\wedge\partial_t \Phi \right] 
\end{equation}
where we have set $a=1$.
We next integrate over the fluctuation fields, that is to ${\bf l}$ 
and $\Delta$. So that we finally obtain the \nls action 
\begin{equation}
S=\int dtdx \left[ \frac{1}{2+\rho}(\partial_t \Phi)^2 
-\frac{1}{2}s^2(2+\frac{\delta\rho}{\delta+\rho})(\partial_x \Phi)^2 
\right]. \end{equation}
The standard parameter of this \nls are
\begin{equation}
g=\frac{1}{s}\sqrt{\frac{2+\rho}{2+\frac{\delta\rho}{\delta+\rho}}} 
\qquad \mbox{and} \qquad
c=s \sqrt{(2+\rho)\left(2+\frac{\delta \rho}{\delta+\rho}\right)}. 
\end{equation}

Now in order to stick to White's notations, we set $\rho=1$ to get~:
\begin{equation}
g=\frac{1}{s}\sqrt{\frac{3(1+\delta)}{2+3\delta}} \qquad 
c=\frac{s}{2}\sqrt{\frac{3(2+3\delta)}{1+\delta}} .
\end{equation}
These are the coupling and the velocity of the effective \nls. 

%%%%%%%%%%%%%%%%%%%%%%%%%%%%%%%%%%%%%%%%%%%%%%%%%%%%%%%%%%%%%%%%%%%%%%
\subsection{Energy gap evaluation in the large $N$-limit} 

A line of reasoning similar to Hida's chain treatment can be applied 
to this \nls. In the large N-component limit, we can derive a saddle 
point equation and obtain the generated mass. The energy gap is then 
given by
\begin{equation}
\Delta_L^s=\Lambda.c.\exp(-\frac{2\pi}{g}). 
\end{equation}
Note that contrary to the Hida's chain case, it is not meaningful to 
stick to the hyperbolic sine function, since 
it is not more precise than the function exponential. With the 
previously computed coupling constant $g$ and velocity $c$, we obtain
\begin{equation}
\Delta_L^s=\Lambda.\frac{s}{2}\sqrt{\frac{3(2+3\delta)}{1+\delta}} 
\exp\left(-2\pi s \sqrt{\frac{2+3\delta}{3(1+\delta)}} \right). 
\end{equation}
Since when $\delta \rightarrow \infty$, we should recover 3/4 of the 
gap $\Delta_C$ of the spin-1 chain, we can rewrite it as~:
\begin{equation}
\frac{\Delta_L^s}{\frac{3}{4}.\Delta_C^s}= 
\sqrt{\frac{2+3\delta}{3(1+\delta)}}
\exp \left[ 2\pi s\left(1-
\sqrt{\frac{2+3\delta}{3(1+\delta)}}\right) \right] . 
\end{equation}

Hence we can relate the spin gap of the antiferromagnetic spin-1 
chain to the spin gap of the antiferromagnetic two-leg spin $1/2$ 
ladder by the formula~:
\begin{equation}
\frac{\Delta_L}{\Delta_C}=
\sqrt{\frac{3}{8}}
\exp \left[ \pi\left( 1-\sqrt{\frac{2}{3}}\right) \right]. 
\end{equation}

With the value $\Delta_C \simeq 0.41$, we obtain the estimate 
$\Delta_L \simeq 0.45$, close to the known\cite{Barnes} value $0.50$.
In figure 5, we have drawn the curve of the normalized spin gap
$\Delta_L/(3/4.\Delta_C)$ w.r.t. $1-2/\pi.\arctan \delta$. We can 
compare the curve with data from White\cite{White}. Not only does 
our result agree qualitatively, but it is also quantitatively quite 
good.

%%%%%%%%%%%%%%%%%%%%%%%%%%%%%%%%%%%%%%%%%%%%%%%%%%%%%%%%%%%%%%%%% 
%%%%%%%%%%%%%%%%%%%%%%%%%%%%%%%%%%%%%%%%%%%%%%%%%%%%%%%%%%%%%%%%% 
\section{Conclusion}
\label{3}

We have constructed non-linear sigma models appropriate to the 
description
of the properties of some generalized spin chains including
ferromagnetic exchanges. In the case of the alternating 
ferro-antiferromagnetic spin chain, our treatment reproduces 
correctly
the smooth crossover from decoupled dimers to the Haldane phase.
This approach may be of relevance to the study of the compound
CuNb$_{2}$O$_{6}$ which is such an alternating chain and has a spin 
gap\cite{Kodama}. 

We have also treated a ladder including ferro bonds so that
the Haldane phase can be reached. Here again there is good
agreement with numerical data.

\acknowledgments

We thank A. V. Chubukov for useful
discussions and collaboration at an early stage of this work.

%%%%%%%%%%%%%%%%%%%%%%%%%%%%%%%%%%%%%%%%%%%%%%%%%%%%%%%%%%%%%%%%% 
%\appendix

%%%%%%%%%%%%%%%%%%%%%%%%%REFERENCES%%%%%%%%%%%%%%%%%%%%%%%%%%%%%% 

\newpage

%%%%%%%%%%%%%%%%%%%%%%%%%%%%%%%%%%%%%%%%%%%%%%%%%%%%%%%%%%%%% 
%%%%%%%%%%%%%%%%%% figures %%%%%%%%%%%%%%%%%%%%%%%%%%%%%%%%%% 

%%%%%%%%%%%%%%%%%%%%%%%%%%%%%%%%%%%%%%%%%%%%%%%% 

	\begin{figure}
	\label{fig1}
	\epsfxsize=10cm	$$ \epsfbox{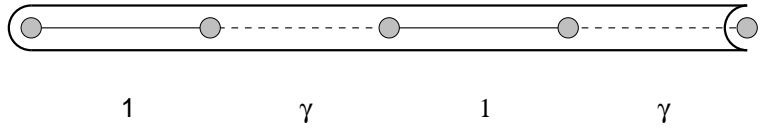} $$	
	\caption{The alternating ferro-antiferromagnetic spin chain. 
	The coupling 
	$\gamma$ is the strength of the ferromagnetic bonds.
	The unit cell needed to construct a \nls contains four spins.}

	\end{figure}

	\begin{figure}
	\label{fig2}
	\epsfxsize=10cm	$$ \epsfbox{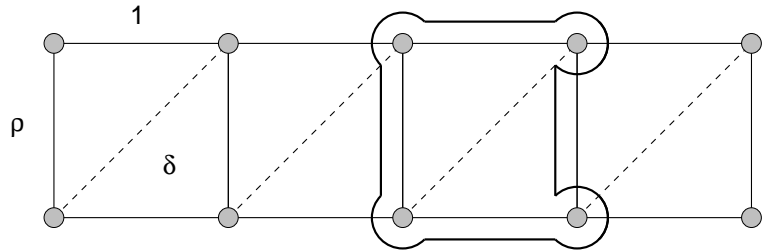} $$	
	\caption{A spin ladder with diagonal couplings $\delta$. The unit
	block to construct a \nls contains again four spins.}

	\end{figure}

	\begin{figure}
	\label{fig3}
	\epsfxsize=10cm	$$ \epsfbox{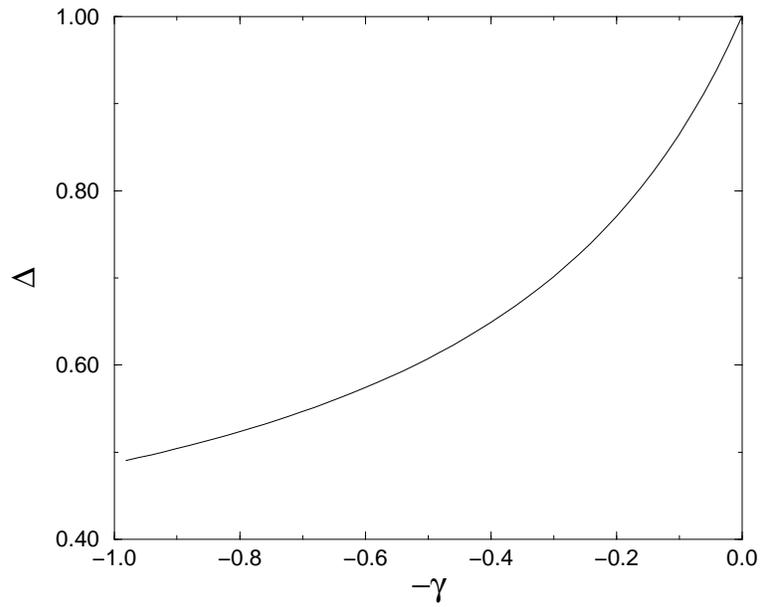} $$	
	\caption{The gap of the alternating chain as a function of 
	$\gamma$.
	The point with decoupled dimers is on the right.}

	\end{figure}

	\begin{figure}
	\label{fig4}
	\epsfxsize=10cm	$$ \epsfbox{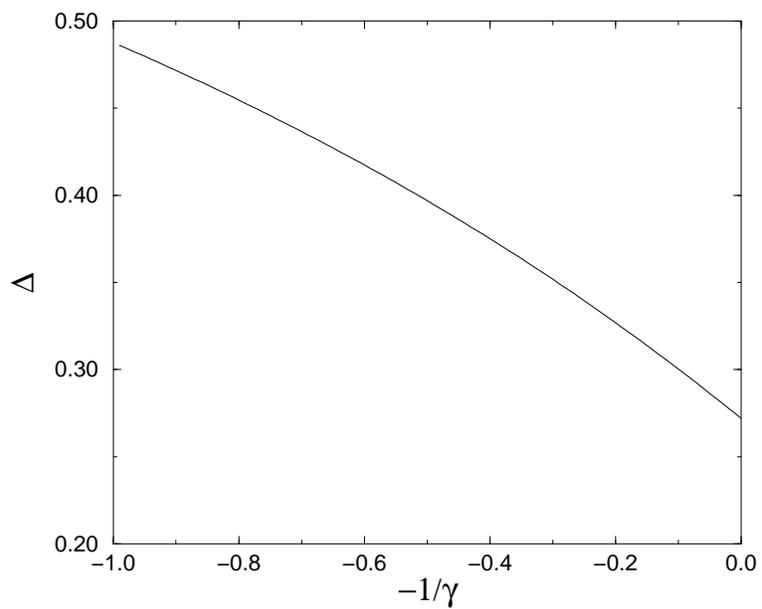} $$	
	\caption{Same as preceding figure but the S=1 chain is recovered
	on the right for $1/\gamma =0$.}

	\end{figure}

	\begin{figure}
	\label{fig5}
	\epsfxsize=10cm	$$ \epsfbox{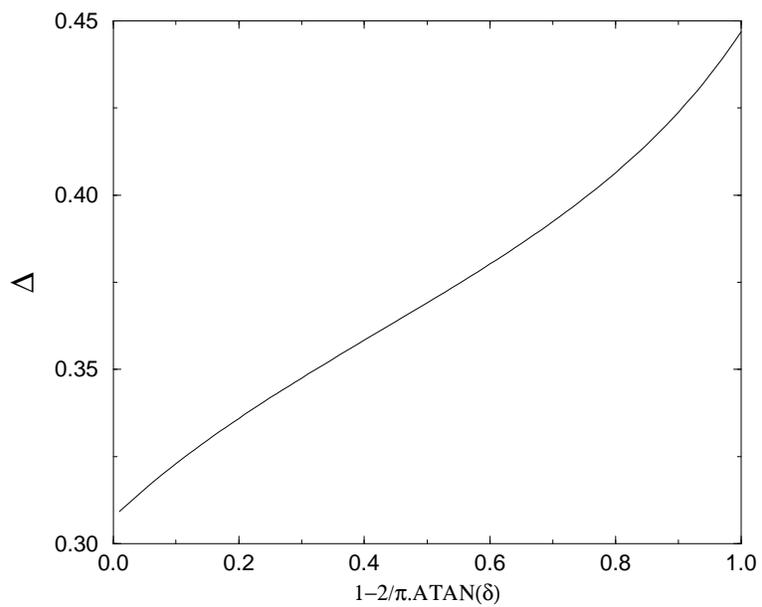}	$$	
	\caption{The gap of the spin ladder with diagonal couplings as
	a function of $\delta$.}

	\end{figure}

%%%%%%%%%%%%%%%%%%%%%%%%%%%%%%%%%%%%%%%%%%%%%%%% 

\end{document}